**Title.**

# A novel method for the evaluation of uncertainty in dose volume histogram computation.

**Authors.**

Francisco Cutanda Henríquez. NW Medical Physics. Christie Hospital. Wilmslow Road. Withington. M20 4BX. Manchester. United Kingdom. Francisco.Cutanda@physics.cr.man.ac.uk.

Silvia Vargas-Castrillón. NW Medical Physics. Christie Hospital. Wilmslow Road. Withington. M20 4BX. Manchester. United Kingdom. Silvia.Vargas@physics.cr.man.ac.uk.

**Abstract.**

Dose volume histograms are a useful tool in state-of-the-art radiotherapy planning, and it is essential to be aware of their limitations. Dose distributions computed by treatment planning systems are affected by several sources of uncertainty such as algorithm limitations, measurement uncertainty in the data used to model the beam and residual differences between measured and computed dose, once the model is optimized. In order to take into account the effect of uncertainty, a probabilistic approach is proposed and a new kind of histogram, a dose-expected volume histogram, is introduced. The expected value of the volume in the region of interest receiving an absorbed dose equal or greater than a certain value is found using the probability distribution of the dose at each point. A rectangular probability distribution is assumed for this point dose, and a relationship is given for practical computations. This method is applied to a set of dose volume histograms for different regions of interest for 6 brain patients, 8 lung patients, 8 pelvis patients and 6 prostate patients planned for IMRT. These results show how dose computation uncertainty has effects on PTV coverage and, to a lesser extent, in dose to organs at risk. This method allows to quantify these effects.

**Introduction.**

Dose volume histograms (DVH) were introduced as a tool for plan evaluation in the late 80s, at a time when three dimensional dose computations were becoming state-of-the-art[1]. Dose volume histograms started being used routinely for plan evaluation and even dose prescription very soon. Currently, dose volume histogram constraints are used as input data for intensity modulated radiation therapy (IMRT) planning. Actually, prescription based on DVHs has replaced the traditional dose to a point approach when prescribing and reporting IMRT treatments, because of the impossibility of fulfilling ICRU requirements[2,3]. Thus, a careful study on accuracy of dose volume histogram computation is now as necessary as for point dose calculations, and most recent reports on treatment planning quality assurance make recommendations about DVH accuracy assessment[4,5,6,7,8,9,10].

Despite these efforts, a statement of uncertainty for a particular DVH value is not as straightforward, as it is with point dose. Since a DVH is a statistical concept, its value for a particular dose level is obtained from many point dose computations, and their uncertainties are not correlated.

Niemierko and Goitein[11], and Lu and Chin[12] published work on DVH computation accuracy, comparing two methods of volume computation: random (or quasi random) sampling and grid placement. Advantages and disadvantages of both methods were the subject of further comments[13,14]. Kooy et al[15] published a new methodology for volume assessment in small, nearly spherical volumes of interest, adapted for its use in radiosurgery plans, which improved volume computation accuracy. The issue of dose uncertainty in DVHs was not addressed in these papers.

Uncertainties for point dose computations arise from several different sources. There is an inherent random (type A) measurement uncertainty in the data used for modeling the beams; there is a type B uncertainty due to algorithm limitations, which depends on patient and organ features (geometry and tissue density) and beam characteristics; and there is also a type B uncertainty due to imperfect matching between measured and computed data when modeling the beam. When specifying dose to a point, a composition of uncertainties from all sources must be used. The problem of combining these uncertainties to obtain the uncertainty of a dose volume histogram value is complex. Different voxels with the same computed absorbed dose can have very different irradiation conditions, and they depend on the beam arrangement and patient anatomy. A probabilistic approach has been chosen in this paper. A modified version of the DVH has been developed, which takes into account the probability of each voxel receiving a dose greater than the dose level considered.

Some applications to clinical plans have been prepared to illustrate the effects of dose uncertainty on DVHs, and a summary of statistical parameters is provided as a result.

**Materials and Methods.**

*Theory.* The standard definition of cumulative dose volume histogram, $DVH_c(x)$, of a region of interest (ROI) at dose level $x$ is: the volume contained in the region of interest receiving a dose equal or greater than $x$[1]. It is common practice to use relative volumes and/or doses, referring volumes to the ROI total volume and doses to an arbitrary chosen level (prescription dose). Another variant of this concept is the differential dose volume histogram, $DVH_d(x)$, which is defined as the volume contained in the region of interest receiving a dose level x. Relationships between both functions can be obtained easily.

$$DVH_c(x) = \int_x^\infty DVH_d(y)dy$$

$$DVH_a(x) = -\frac{d}{dx}DVH_c(x)$$

Dose-volume histograms are usually computed for a discreet set of dose intervals of uniform length. Random sampling or regular grids are possible methods to sample dose points inside the region of interest. In practice, dose-volume histograms are computed on a discreet set of $N$ voxels with volume $v_i$, where the computed dose is assumed to be approximately constant and equal to the dose at a representative point $z_i$ (its centre).

Because of uncertainty, absorbed dose at sample point $z_i$ can be represented mathematically as a random variable $\Delta_i$, with density function $f_i(\delta_i)$, whose mean is $D(z_i)$ (computed dose in $z_i$), and its variance $\sigma_i^2$. The knowledge about this dose distribution is usually its variance or standard deviation. Standard deviations are the recommended parameter to evaluate uncertainties, because in this way uncertainties from different sources can be combined according to certain rules[16,17,18], regardless of the particular distribution of probability.

We are interested in evaluating probabilities for this random variable $\Delta_i$. Depending on the sort of dose uncertainty, and whether it is a type A or type B uncertainty, different probability density functions could be assumed. If there is a dominant type A component, due to experimental uncertainty, or to a composition of many small uncertainty sources, then the Central Limit Theorem may be applicable and a Gaussian distribution can be used.

$$f_i(\delta_i) = \frac{1}{\sqrt{2\pi}\sigma_i} e^{-\frac{(\delta_i - D(z_i))^2}{\sigma_i^2}}$$

In other situations, there is no information about which the probability distribution for our random variable is, and a rectangular density function is assumed, giving equal probability to any result within an upper and a lower bound (ISO Guide[16]).

Regardless of the choice, it is assumed that the standard deviation is a constant percentage of the computed dose, i.e., relative uncertainty is constant within the region of interest R, so $\sigma_y = u \cdot D(y)$ (see below). We can then define a density function $f(w)$ verifying

$$f_i(\delta_i) = f\left(\frac{\delta_i - D(z_i)}{\sigma_i}\right) = f\left(\frac{\delta_i - D(z_i)}{u \cdot D(z_i)}\right)$$

and $f(w)$ has mean 0 and standard deviation 1.

Cumulative dose-expected volume histogram, $DeVH_c(x)$, for the region of interest R and dose level x is defined as the expected value of the volume contained in R receiving a dose equal or greater than x.

If $T_i^x$ is defined as a random variable with value 1 when $\delta_i \geq x$ and 0 otherwise, then the sum $\sum_{i=1}^{N} T_i^x \cdot v_i$ is a random variable corresponding to the volume receiving a dose greater than x. Each voxel adds the value $v_i$ to this summation according to $T_i^x$. The dose expected volume histogram is obtained as the expectation value of this sum:

$$DeVH_c(x) = E\left[\sum_{i=1}^{N} T_i^x \cdot v_i\right] = \sum_{i=1}^{N} E[T_i^x] \cdot v_i$$

$T_i^x$ is a Bernoulli random variable with parameter $p = P[\delta_i \geq x]$, mean $p$ and variance $p(1-p)$:

$$E[T_i^x] = P[\delta_i \geq x] = \int_x^\infty f\left(\frac{t - D(z_i)}{u \cdot D(z_i)}\right) \cdot dt = \left[1 - F\left(\frac{x - D(z_i)}{u \cdot D(z_i)}\right)\right]$$

$F(s) = \int_{-\infty}^{s} f(w) \cdot dw$ is the distribution function associated to $f(w)$. Thus, we obtain the following equation for $DeVH_c(x)$:

$$DeVH_c(x) = \sum_{i=1}^{N}\left[1 - F\left(\frac{x - D(z_i)}{u \cdot D(z_i)}\right)\right] \cdot v_i$$

Each voxel in the region of interest $R$ is weighted according to the probability tail to the right side of $x$. Each term in the summation depends on $z_i$ through $D(z_i)$. Thus, adding up all voxels with computed dose between $d_j$ and $d_{j+1}$, corresponding to the $j$-eth dose bin, the final equation for $DeVH_c(d_j)$ is

$$DeVH_c(d_j) = \sum_{k=1}^{\infty}\left[1 - F\left(\frac{d_j - d_k}{u \cdot d_k}\right)\right] \cdot DVH_d(d_k) \qquad (1)$$

We have used the fact that the sum of all $v_i$ corresponding to voxels with computed dose between $d_k$ and $d_{k+1}$ is $DVH_d(d_k)$. There is a finite number of terms in the summation, because there is a maximum dose.

This equation is consistent with the usual $DVH$ definition. If a Dirac delta is used as probability distribution (no dose uncertainty), then $F(w)$ is the Heavyside step function at $0$.

$$F(w) = H(w) = \begin{cases} 0 & w < 0 \\ 1 & w \geq 0 \end{cases}$$

and the following equation

$$DeVH_c(d_j) = \sum_{k=1}^{\infty}\left[1 - H\left(\frac{d_j - d_k}{u \cdot d_k}\right)\right] \cdot DVH_d(d_k) = \sum_{k=j}^{\infty} DVH_d(d_k) = DVH_c(d_j)$$

is obtained.

Now that $DeVH_c$ has been obtained as a summation in dose, it is possible to take into account differences in relative dose uncertainty throughout the region of interest, and relax the assumption made above. Accuracy of dose computation in low dose, low gradient regions is considered worse than in points inside the beam[19,20]. This fact could be partially taken into account by splitting the summation in different parts, using different $u$ values for each range of computed doses. This approach has not been used in this paper, and a single value is used for $u$, which is the uncertainty for high dose points. Its use can only underestimate the effect of uncertainty.

An upper bound for the uncertainty in $DeVH_c$ values can be obtained adding variances of $T_i^x$. These random variables have correlation: errors arising from algorithm limitations in adjacent voxels, for instance, are likely to be similar, although this correlation is lost for distant voxels. Thus, uncertainty should be less than the value obtained from the sum of squares because of correlation terms.

$$u(DeVH_c(d_j))^2 \leq \sum_{k=1}^{\infty} \left[1 - F\left(\frac{d_j - d_k}{u \cdot d_k}\right)\right] \cdot F\left(\frac{d_j - d_k}{u \cdot d_k}\right) \cdot DVH_d(d_k)$$

At this point, a choice has to be made about the distribution associated to point dose computations. Having a non negligible Type B component arising from algorithm inaccuracy and limitations in beam modeling a rectangular distribution is the recommended option according to ISO[16]. Its density function (with mean 0 and variance 1) is

$$f(w) = \begin{cases} \dfrac{1}{2\sqrt{3}} & w \in \left[-\sqrt{3}, \sqrt{3}\right] \\ 0 & elsewhere \end{cases}$$

And its distribution function is

$$F(w) = \begin{cases} 0 & w < -\sqrt{3} \\ \dfrac{w + \sqrt{3}}{2\sqrt{3}} & -\sqrt{3} \leq w < \sqrt{3} \\ 1 & \sqrt{3} \leq w \end{cases}$$

Let us derive $DeVH_c$ for this case.

$$1 - F\left(\frac{d_j - d_k}{u \cdot d_k}\right) = \begin{cases} 0 & d_k \leq \dfrac{d_j}{1 + \sqrt{3} \cdot u} \\ \dfrac{d_j - d_k(1 - \sqrt{3}u)}{2\sqrt{3}ud_k} & \dfrac{d_j}{1 + \sqrt{3} \cdot u} < d_k \leq \dfrac{d_j}{1 - \sqrt{3} \cdot u} \\ 1 & \dfrac{d_j}{1 - \sqrt{3} \cdot u} < d_k \end{cases}$$

And the dose expected volume histogram is

$$DeVH_c(d_j) = \sum_{d_k \in \left(\frac{d_j}{1+\sqrt{3}\cdot u}, \frac{d_j}{1-\sqrt{3}\cdot u}\right]} \frac{d_j - d_k(1-\sqrt{3}u)}{2\sqrt{3}ud_k} \cdot DVH_d(d_k) + \sum_{d_k > \frac{d_j}{1-\sqrt{3}\cdot u}} DVH_d(d_k) =$$

$$= \sum_{d_k \in \left(\frac{d_j}{1+\sqrt{3}\cdot u}, \frac{d_j}{1-\sqrt{3}\cdot u}\right]} \frac{d_j - d_k(1-\sqrt{3}u)}{2\sqrt{3}ud_k} \cdot DVH_d(d_k) + DVH_c\left(\frac{d_j}{1-\sqrt{3}\cdot u}\right) \quad (2)$$

It is worth comparing this equation with the following one,

$$DVH_c(d_j) = \sum_{d_k > d_j} DVH_d(d_k) = \sum_{d_k \in \left(d_j, \frac{d_j}{1-\sqrt{3}\cdot u}\right]} DVH_d(d_k) + DHV_c\left(\frac{d_j}{1-\sqrt{3}\cdot u}\right)$$

It can be observed that there is a redistribution of dose bins contributing to the value at level $d_j$ in the cumulative dose expected volume histogram. A certain amount of volume is now added from bins below $d_j$, which did not contribute for $DVH_c(d_j)$, and some volume is removed from higher dose bins, being the dose bins involved in this rearrangement the ones in the interval $\left(\frac{d_j}{1+\sqrt{3}\cdot u}, \frac{d_j}{1-\sqrt{3}\cdot u}\right]$. Figure 1 helps to illustrate this rearrangement.

The uncertainty (standard deviation) of $DeVH_c(x)$ is

$$u(DeVH_c(d_j)) \leq \left\{ \sum_{d_k \in \left(\frac{d_j}{1+\sqrt{3}\cdot u}, \frac{d_j}{1-\sqrt{3}\cdot u}\right]} \frac{1}{4} \cdot \left(1 - \frac{(d_j - d_k)^2}{3u^2 d_k^2}\right) \cdot DVH_d(d_k) \right\}^{\frac{1}{2}}$$

Larger differences between $DeVH(d_j)$ and $DVH(d_j)$ can be expected when the differential dose volume histogram has large values at and around $d_j$. In other words, the worst case scenario would occur when the cumulative dose volume histogram is rapidly decreasing at $d_j$. Obviously, a large uncertainty value gives also larger differences.

*Practical application.* Equation 2 has been used to obtain dose-expected volume histograms for a set of plans and regions of interest: eight standard and six IMRT prostate plans for PTV, rectum and bladder; eight lung plans for PTV, both lungs and spinal cord; six brain plans for PTV and brain stem. Every plan has been the result of standard planning procedures in our institution. The treatment planning system used is Pinnacle v7.6g (Philips. Milpitas. CA). DVH dose bins were 25 to 32 cGy wide, which means that there were around 200 points in each histogram. Dose and volume have always been used in absolute values (in cGy and cm$^3$).

A 3.6% value of uncertainty has been considered for every region of interest. This choice is based on the results given for Pinnacle by Venselaar and Welleweerd[21] in their survey of treatment planning systems in the Netherlands. The confidence level for the "irregular block case" studied by them has been chosen, because MLC shaped beams have been used in every patient. This confidence level is 3.3% for Pinnacle. IAEA TRS-398[22] provides a standard uncertainty value of 1.5% (IAEA) for absolute dose determination in reference conditions. Both values are standard deviations and its composition gives the value 3.6 % that has been used in this paper. This value is an underestimation, because tests in this survey are still ideal situations where only one feature is taken into account at a time, and the IAEA value applies only to reference conditions. The same authors propose higher tolerance levels for combinations of irregular beams, in inhomogeneous media, in other publication[20], and they can be used as an alternative. Different planning systems or different kind of treatments could lead to other figures for the uncertainty.

For every region of interest several parameters were computed from their corresponding DVHs and DeVHs: mean dose, standard deviation, maximum dose (defined as the dose corresponding to 1 cm$^3$ in a cumulative histogram), minimum dose, median, first and third quartile. The differences in these parameters between DVHs and DeVHs are given as quantitative results. Some DVH values are studied also, and their value with uncertainty is computed from DeVH. A number of dose-volume parameters has been proved to have predictive value for certain side effects, according to the literature[23,24,25,26,27]: V20 and V30 for lungs, V65 for bladder, V60 or V72 for rectum in IMRT prostate treatments. Special fractionation schemes for prostate treatments have been developed in our institution, being the standard for conventional conformal radiotherapy 50 Gy in 16 fractions in 22 days[28], and a simultaneous boost technique for IMRT treatments with 54 Gy (PTV1), 57 Gy (PTV2) and 60 Gy (PTV3) in 3 Gy fractions[29]. As examples of application of DeVHs, values for V40 for bladder and rectum in conventional prostate treatments are given, and V50 in IMRT treatments. V50 for brain stem in brain treatments and V30 for lung are also studied as examples. The volume of the 95% isodose, a coverage descriptor for PTVs, has also been obtained. These magnitudes are shown as possible applications of DeVHs and do not reflect current practice in our department.

**Results and discussion**

Figure 2 illustrates what the impact of uncertainty in DeVH is. Several values of uncertainty have been used. We can see that the effect is more remarkable when the DVH gradient is higher. Thus, PTVs are more sensitive to uncertainty than large volume organs at risk like lungs. Figures 3-5 show a comparison of DVHs and DeVHs, for several organs. It shows what the effect is depending on the organ that is being considered. A common feature in all DeVHs is the increase in maximum dose.

Tables 1-4 show the differences in each of the parameters between DeVH and DVH for each region of interest for each of the treatments studied. Although the effect in mean dose is limited, maximum and minimum doses are further from these mean doses, making the dose distribution more inhomogeneous. The same feature is apparent when considering median dose and first and third quartiles.

Table 5 shows values of V30, V40, V50, V65, V60, V95 depending on the region of interest, computed with DVH and DeVH, and its uncertainty in the latter case. V95 is shown as percentage of the prescribed dose. It can be seen that the 95% isodose has a lower volume if uncertainty is taken into account. The plan should be corrected if tumour coverage is to be insured despite uncertainty.

**Conclusions.**

Absorbed dose measurements and computations have some degree of uncertainty due to multiple causes. The figure of 5% (expanded uncertainty equivalent to twice a standard uncertainty) given in ICRU Report No. 24[30] is often quoted as the goal for radiotherapy delivery needed to ensure tumour eradication. It is also widely recognized that this is a difficult goal to achieve, and that, although commercial computation algorithms are improving, treatment planning systems contribute to this overall uncertainty.

Dose expected volume histograms are a useful tool when uncertainty has to be taken into account for decision making. Results show that dose homogeneity could not be as good as the treatment planning system shows for PTVs, and, if maximum dose is a concern, it is higher than the value given by the treatment planning system. If a goal has to be achieved safely, some margin for uncertainty has to be considered.

We are used to attach an uncertainty estimation to point dose computations, for instance, ICRU-50 point dose values. But measurement and computation uncertainty has effect on every point in our patient, and can compromise PTV coverage. The difference between conventional DVHs and DeVHs shows how large this influence can be. Dose expected volume histograms are a suitable tool for taking these effects into account, and their implementation into routine practice is easy. Users can make their own choices as to what probability distribution has to be used, how large the dose uncertainty is in their case, and what conclusions, or actions, can be reached. This way, it becomes possible to report external beam radiotherapy with a description of uncertainty, either when prescription is based on dose to a point or when it is based on dose volume relationships.

**FIGURES.**

**Figure 1.** Contribution of the different dose bins to DeVH($d_j$). For each dose bin the rectangular probability distribution is shown, with the area contributing to DeVH($d_j$) in grey color. The sum of the grey areas, weighted by the differential DVH, equals DeVH($d_j$).

**Figure 2.** DVH and DeVH for the PTV of an IMRT prostated plan. Five DeVH have been superimposed, with different dose uncertainty values: 1%, 2%, 3%, 4%, 5%.

**Figure 3.** DVH and DeVH (with u=3.6%) for a prostate treatment planned with step and shoot IMRT. a) PTV, b) Bladder, c) Rectum.

**Figure 4.** DVH and DeVH (with u=3.6%) for a lung treatment. a) PTV, b) Spinal cord, c) Left lung.

**Figure 5.** DVH and DeVH (with u=3.6%) for a brain treatment. a) PTV, b) Brain stem.

**Table 1.** Mean dose, standard deviation, maximum dose, minimum dose, median dose, first and third quartiles computed with DeVH and DVH for conventional prostate plans. The values shown are the ones for the case with largest differences for each region of interest. All values in the table are in cGy.

| Prostate | PTV | | Bladder | | Rectum | |
|---|---|---|---|---|---|---|
| | DVH | DeVH | DVH | DeVH | DVH | DeVH |
| Mean Dose | 4919.7 | 4907.12 | 4271.5 | 4258.9 | 3681.4 | 3668.8 |
| Standard deviation | 2.9 | 11.9 | 55.6 | 56.5 | 56.3 | 57.0 |
| Maximum Dose | 4987.5 | 5261.4 | 4968.9 | 5214.9 | 4919.5 | 5103.7 |
| Minimum Dose | 4725.0 | 4425.0 | 2600.0 | 2425.0 | 2725.0 | 2550.0 |
| Median Dose | 4940.7 | 4919.3 | 4826.3 | 4683.1 | 3461.5 | 3451.0 |
| First quartile | 4908.4 | 4765.9 | 3510.3 | 3489.8 | 2811.1 | 2844.5 |
| Third quartile | 4964.9 | 5072.7 | 4935.2 | 4932.9 | 4664.6 | 4590.8 |

**Table 2.** Mean dose, standard deviation, maximum dose, minimum dose, median dose, first and third quartiles computed with DeVH and DVH for IMRT prostate plans. The values shown are the ones for the case with largest differences for each region of interest. All values in the table are in cGy.

| Prostate IMRT | PTV1 | | PTV2 | | PTV3 | | Bladder | | Rectum | |
|---|---|---|---|---|---|---|---|---|---|---|
| | DVH | DeVH | DVH | DeVH | DVH | DeVH | DVH | DeVH | DVH | DeVH |
| Mean Dose | 6912.7 | 6893.7 | 7175.2 | 7156.1 | 7338.1 | 7319.0 | 5136.5 | 5117.5 | 4925.2 | 4906.2 |
| Standard deviation | 27.4 | 31.9 | 15.0 | 22.6 | 9.3 | 19.6 | 99.6 | 100.4 | 92.9 | 93.7 |
| Maximum Dose | 7600.0 | 8056.0 | 7600.0 | 8056.0 | 7600.0 | 8056.0 | 7600.0 | 8056.0 | 7334.0 | 7790.0 |
| Minimum Dose | 5320.0 | 5016.0 | 6384.0 | 5966.0 | 6954.0 | 6498.0 | 1254.0 | 1140.0 | 684.0 | 608.0 |
| Median Dose | 6998.0 | 6932.6 | 7203.8 | 7170.1 | 7375.6 | 7335.2 | 5232.8 | 5204.0 | 4824.0 | 4817.4 |
| First quartile | 6597.0 | 6567.1 | 7028.5 | 6927.5 | 7245.9 | 7106.5 | 3751.6 | 3758.4 | 4135.2 | 4118.5 |
| Third quartile | 7277.1 | 7275.7 | 7384.5 | 7426.4 | 7479.3 | 7569.4 | 6613.0 | 6520.1 | 6183.5 | 6143.5 |

**Table 3.** Mean dose, standard deviation, maximum dose, minimum dose, median dose, first and third quartiles computed with DeVH and DVH for lung plans. The values shown are the ones for the case with largest differences for each region of interest. All values in the table are in cGy.

| Lung | PTV | | Spinal Cord | | Left Lung | | Right Lung | |
|---|---|---|---|---|---|---|---|---|
| | DVH | DeVH | DVH | DeVH | DVH | DeVH | DVH | DeVH |
| Mean Dose | 5414.4 | 5399.7 | 822.3 | 823.4 | 406.7 | 394.5 | 2238.9 | 2225.4 |
| Standard deviation | 10.3 | 16.4 | 51.3 | 51.3 | 30.4 | 30.5 | 126.7 | 126.9 |
| Maximum Dose | 5830.7 | 6182.3 | 2402.6 | 2549.1 | 1699.4 | 1787.3 | 5830.7 | 6182.3 |
| Minimum Dose | 4336.4 | 4043.4 | 0.0 | 0.0 | 0.0 | 0.0 | 0.0 | 0.0 |
| Median Dose | 5428.2 | 5413.4 | 347.0 | 344.9 | 129.3 | 116.4 | 1921.6 | 1908.1 |
| First quartile | 5340.9 | 5232.1 | 153.0 | 152.2 | 69.1 | 55.1 | 308.5 | 294.6 |
| Third quartile | 5528.2 | 5596.6 | 1733.7 | 1720.0 | 950.7 | 931.1 | 3736.6 | 3750.7 |

**Table 4.** Mean dose, standard deviation, maximum dose, minimum dose, median dose, first and third quartiles computed with DeVH and DVH for brain plans. The values shown are the ones for the case with largest differences for each region of interest. All values in the table are in cGy.

| Brain | PTV | | Brain stem | |
|---|---|---|---|---|
| | DVH | DeVH | DVH | DeVH |
| Mean Dose | 6009.7 | 5994.2 | 2573.5 | 2558.0 |
| Standard deviation | 7.9 | 16.2 | 76.6 | 76.9 |
| Maximum Dose | 6231.0 | 6603.0 | 5580.0 | 5921.0 |
| Minimum Dose | 3658.0 | 3410.0 | 279.0 | 248.0 |
| Median Dose | 6035.8 | 6012.6 | 2418.7 | 2408.3 |
| First quartile | 5993.1 | 5822.4 | 2002.5 | 1976.2 |
| Third quartile | 6091.5 | 6207.1 | 3576.4 | 3544.4 |

**Table 5.** Values for different parameters computed from DVH and DeVH. For each parameter, values are shown for the patients for which DVH and DeVH give the largest and the smallest difference.

| Region of interest | | Minimum difference case | | Maximum difference case | |
|---|---|---|---|---|---|
| | | DVH | DeVH | DVH | DeVH |
| Prostate | PTV V95 (%) | 99.95 | 81.1 ± 3.0 | 99.8 | 72.9 ± 3.5 |
| | Bladder V40 (cm$^3$) | 34.95 | 34.91 ± 0.59 | 46.8 | 45.3 ± 1.2 |
| | Rectum V40 (cm$^3$) | 19.47 | 19.30 ± 0.88 | 23.61 | 23.42 ± 0.94 |
| Prostate IMRT | PTV V95 (%) | 100.0 | 88.9 ± 2.6 | 99.6 | 80.6 ± 3.2 |
| | Bladder V50 (cm$^3$) | 32.8 | 32.2 ± 1.5 | 26.0 | 24.2 ± 1.8 |
| | Rectum V50 (cm$^3$) | 25.3 | 25.2 ± 1.2 | 27.6 | 26.7 ± 1.7 |
| Lung | PTV V95 (%) | 96.3 | 79.4 ± 2.3 | 88.0 | 82.0 ± 1.1 |
| | Left lung V30 (cm$^3$) | 61.00 | 61.3 ± 2.7 | 330.5 | 328.6 ± 2.5 |
| | Right lung V30 (cm$^3$) | 182.6 | 179.8 ± 3.1 | 1102.82 | 1094.23 ± 4.5 |
| Brain | PTV V95 (%) | 97.7 | 90.0 ± 5.6 | 99.9 | 83.2 ± 3.0 |
| | Brain stem V50 (cm$^3$) | 0.27 | 0.28 ± 0.09 | 9.5 | 9.3 ± 1.3 |

**Figure 1**

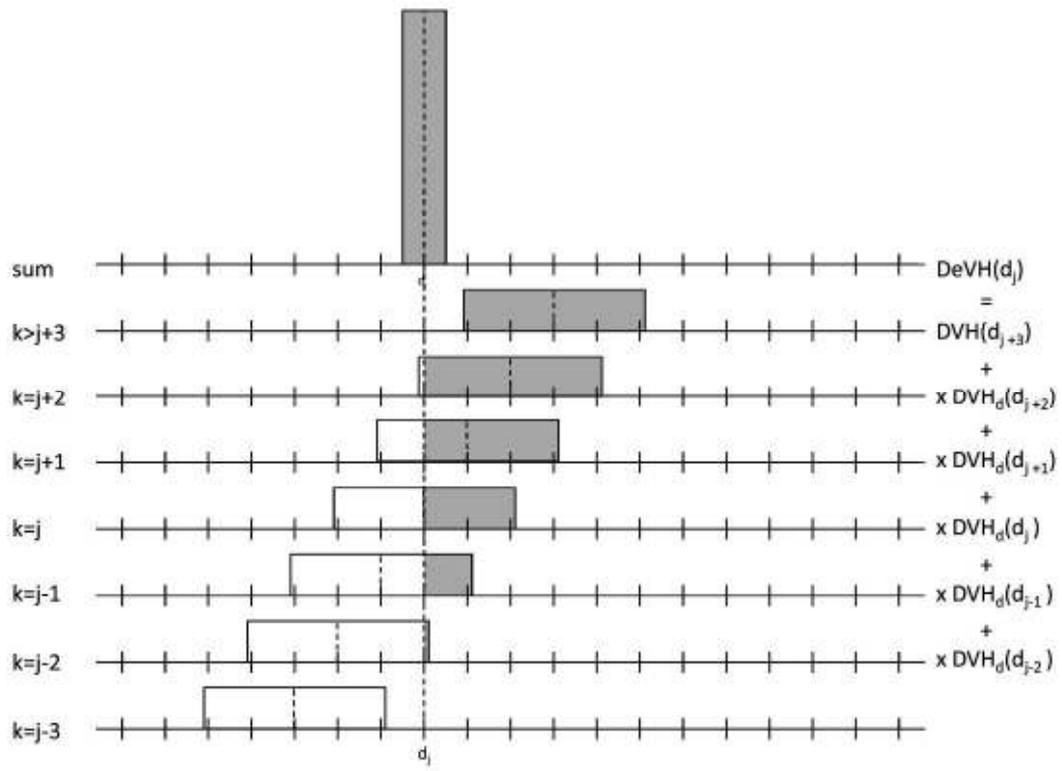

**Figure 2**

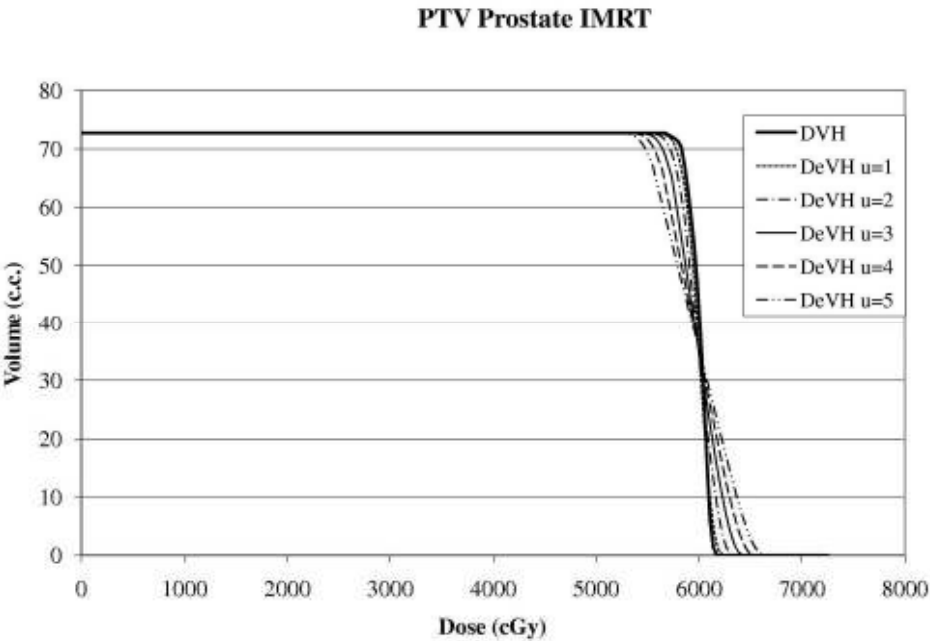

**Figure 3a**

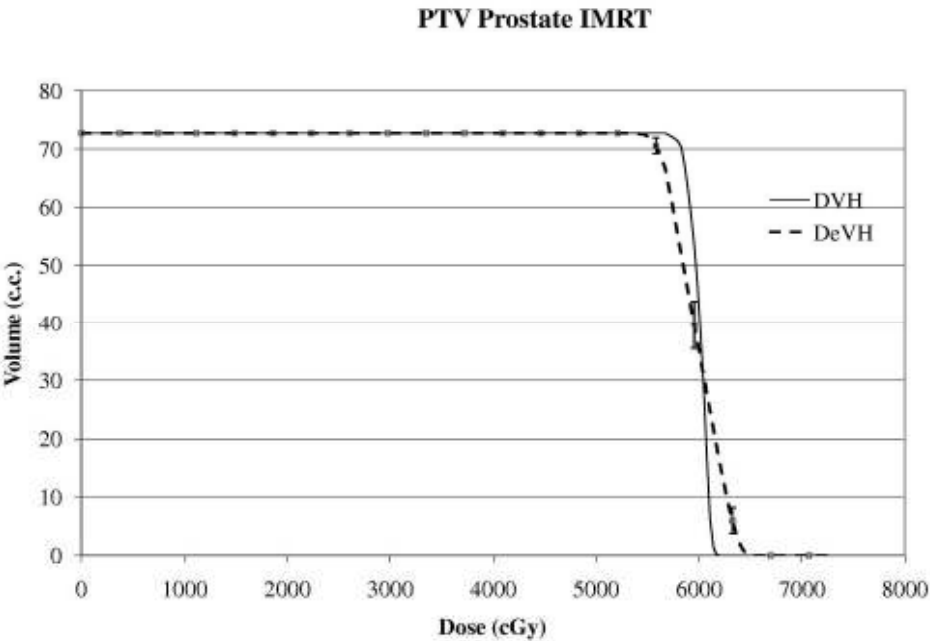

**Figure 3b**

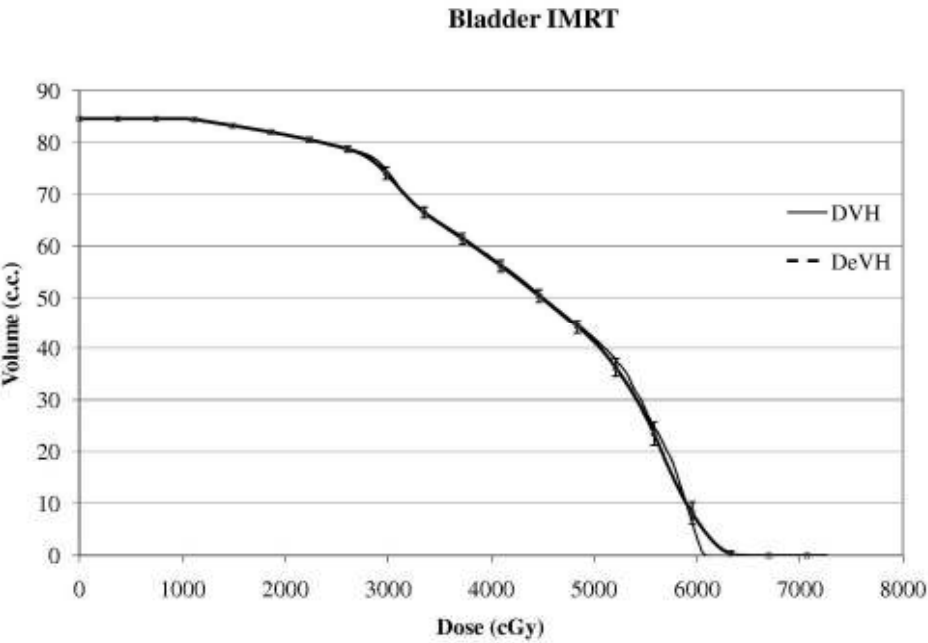

**Figure 3c**

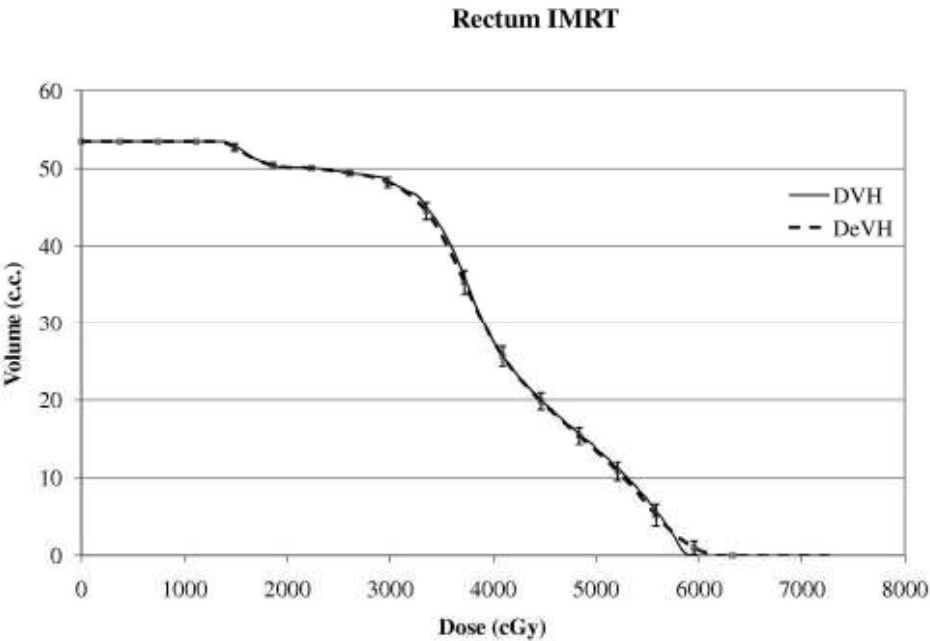

**Figure 4a**

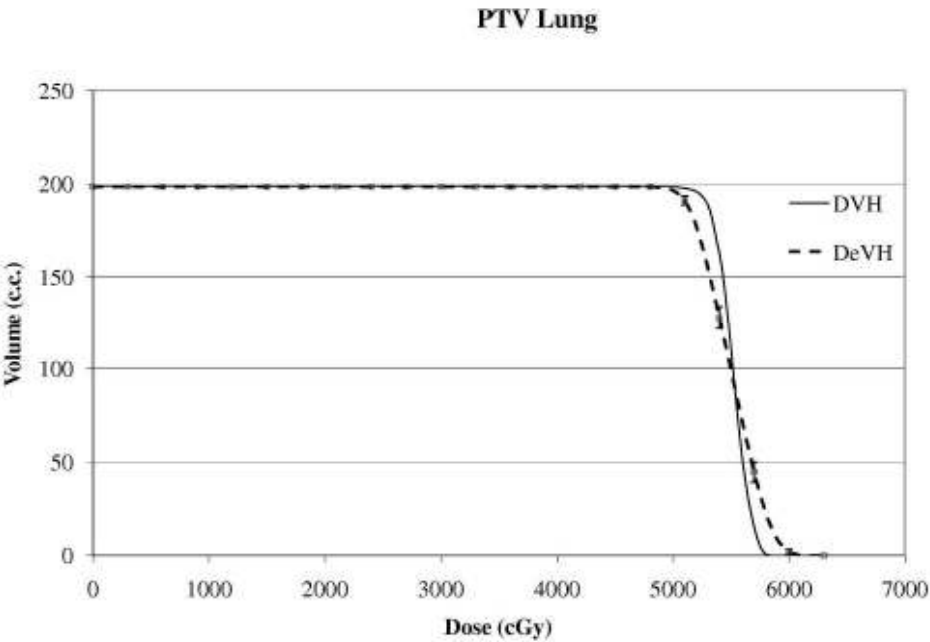

**Figure 4b**

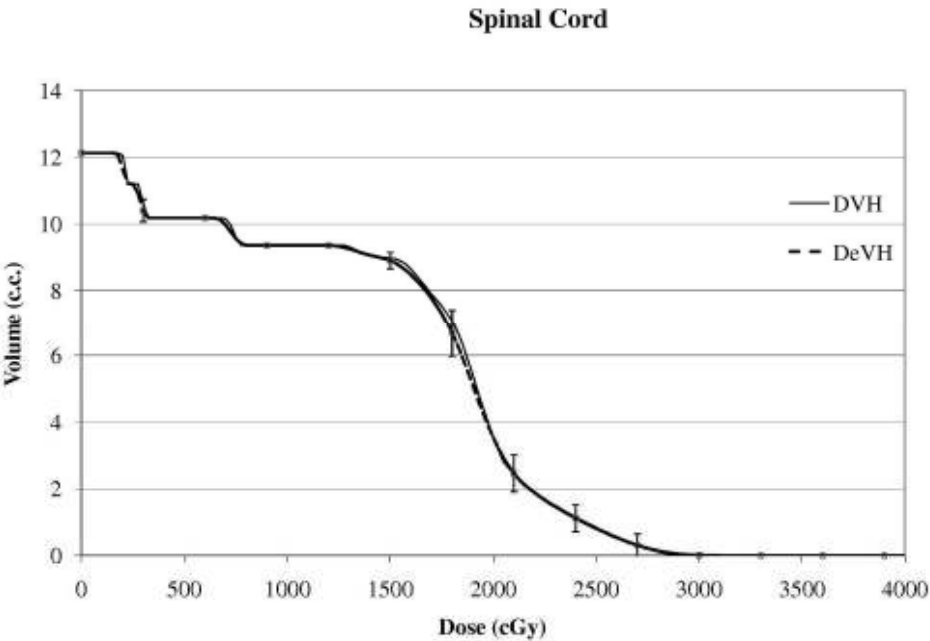

**Figure 4c**

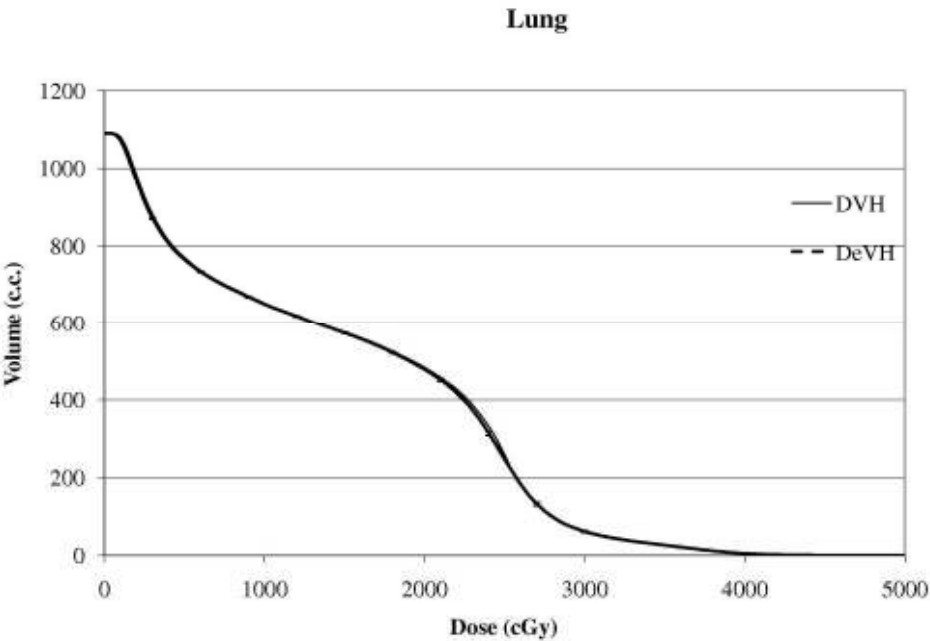

**Figure 5a**

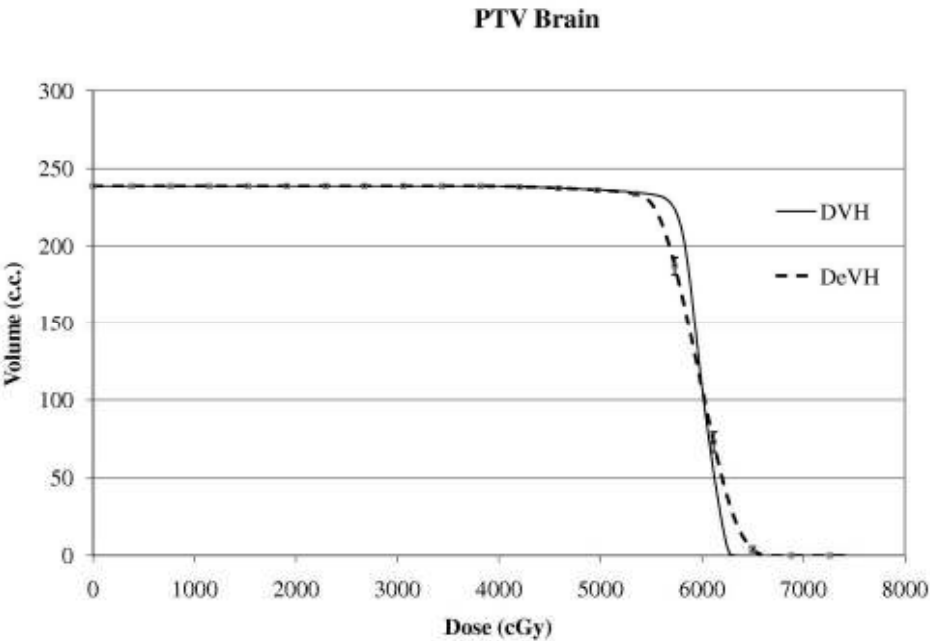

**Figure 5b**

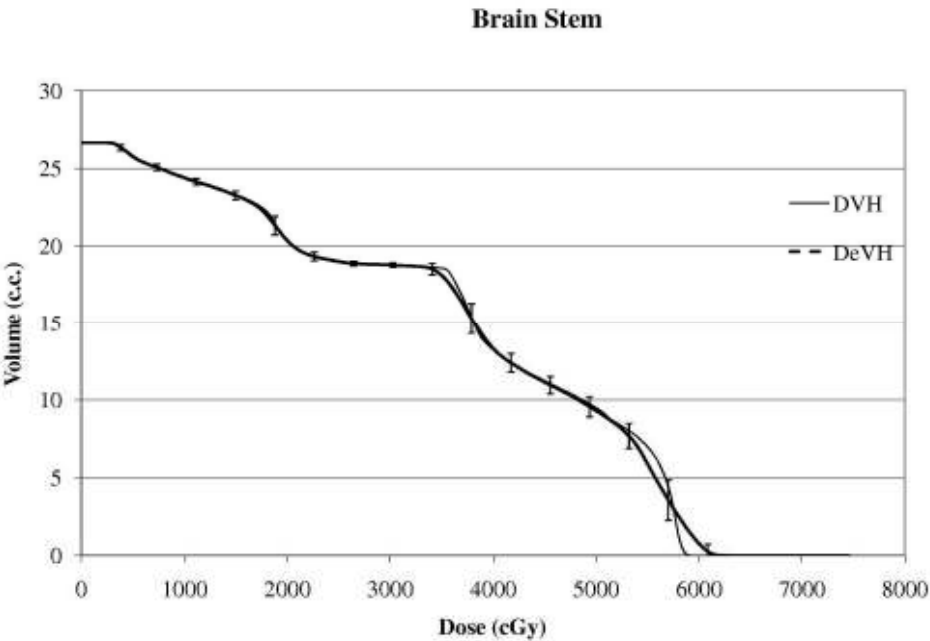